\documentclass{article}%
\usepackage{amsfonts}
\usepackage{graphicx}
\usepackage{amsmath}
\usepackage{amssymb}%
\providecommand{\U}[1]{\protect\rule{.1in}{.1in}}
\newtheorem {theorem}{Theorem}[section]
\newtheorem {proposition}{Proposition}[section]

\newtheorem{lemma}{Lemma}[section]

\newtheorem{remark}{Remark}[section]

\begin{document}

\begin{center}
{\LARGE Asymptotic Properties of Self-Normalized Linear Processes with Long
Memory }\vskip15pt

\bigskip Magda Peligrad$^{a}$\footnote{Supported in part by a Charles Phelps
Taft Memorial Fund grant and NSA\ grant H98230-09-1-0005.} and Hailin
Sang$^{b}$
\end{center}

$^{a}$ Department of Mathematical Sciences, University of Cincinnati, PO Box
210025, Cincinnati, OH 45221-0025, USA. E-mail address: peligrm@ucmail.uc.edu

\bigskip

$^{b}$ National Institute of Statistical Sciences, PO Box 14006, Research
Triangle Park, NC 27709, USA. E-mail address: sang@niss.org

\begin{center}
\bigskip
\end{center}



\bigskip

\textbf{Abstract}

\bigskip

In this paper we study the convergence to fractional Brownian motion for long
memory time series having independent innovations with infinite second moment.
For the sake of applications we derive the self-normalized version of this
theorem. The study is motivated by models arising in economical applications
where often the linear processes have long memory, and the innovations have
heavy tails.

\section{Introduction and notations}

In this paper we study the asymptotic properties of a causal linear process
\begin{equation}
X_{k}=\sum_{i\geq0}a_{i}\varepsilon_{k-i}\label{X}%
\end{equation}
when the i.i.d. innovations $\{\varepsilon,$ $\varepsilon_{n};n\in
\mathbb{Z}\}$ have infinite variance and $\{a_{i};i\geq0\}$ is a sequence of
real constants such that $X_{k}$ is well defined. More precisely, everywhere
in the paper, we assume that the innovations are centered and in the domain of
attraction of a normal law. This means that the variables are independent,
identically distributed,
\begin{equation}
E\varepsilon=0\label{centered}%
\end{equation}
and
\begin{equation}
l(x)=E\varepsilon^{2}I(|\varepsilon|\leq x)\text{ is a slowly varying function
at }\infty\text{ .}\label{DA}%
\end{equation}
We say that $h(t),$ defined for $t\geq0$, is slowly varying if it is positive
and measurable on $[A,\infty),$ for some $A>0,$ and if for any $\lambda>0$, we
have $\lim_{x\rightarrow\infty}h(\lambda x)/h(x)=1$ (Seneta, 1970, Definition 1.1).

We define
\[
S_{n}=\sum_{i=1}^{n}X_{i}\text{ .}%
\]
The central limit theorem for $S_{n}$ with i.i.d. innovations and infinite
variance when $\sum_{i\geq0}|a_{i}|<\infty\ $ was studied by many authors. We
mention among them, Knight (1991), Mikosch et al (1995), Wu (2003). For this
case the central limit theorem was obtained under a normalization that is
regularly varying with exponent $1/2.$

The purpose of this paper is to investigate the central limit theorem in its
functional form for the case when
\begin{equation}
a_{n}=n^{-\alpha}L(n),\text{ where }1/2<\alpha<1, n\geq1 \label{dcoeff}%
\end{equation}
and $L(n)$ is a slowly varying function at $\infty$ in the strong sense (i.e.
there is a slowly varying function $h(t)$ such that $L(n)=h(n)$). Notice that,
by the definition of slowly varying function, the coefficients $a_{n}$ are
positive for $n$ sufficiently large. We shall obtain convergence in
distribution under a normalization that is regularly varying with exponent
$3/2-\alpha$ which is strictly larger than $1/2.$ This is the reason why the
time series we consider has long memory.

To give an example of a linear process of this type we mention the
fractionally integrated processes since they play an important role in
financial time series modeling and they are widely studied. Such processes are
defined for $0<d<1/2$ by
\begin{equation}
X_{k}=(1-B)^{-d}\varepsilon_{k}=\sum_{i\geq0}a_{i}\varepsilon_{k-i}\text{ with
}a_{i}=\frac{\Gamma(i+d)}{\Gamma(d)\Gamma(i+1)}\,\text{\ ,}
\label{deffractlin}%
\end{equation}
where $B$ is the backward shift operator, $B\varepsilon_{k}=\varepsilon_{k-1}%
$. For this example, by the well known fact that for any real $x,$
$\lim_{n\rightarrow\infty}\Gamma(n+x)/n^{x}\Gamma(n)=1,$ we have$\ \lim
_{n\rightarrow\infty}a_{n}/n^{d-1}=1/\Gamma(d)$.

The CLT\ in its functional form was intensively studied for the case of i.i.d.
innovations with finite second moment. We refer to Davydov (1970), Taqqu
(1975), Phillips and Solo (1992), Wang et al (2003), Wu and Min (2005),
Dedecker et al (2009), among others. Invariance principles (or functional
central limit theorems) play an important role in econometrics and statistics.
For example, to obtain asymptotic distributions of unit-root test statistics,
researchers have applied invariance principles of various forms; see Phillips
(1987) and Wu (2006).

We shall derive here the central limit theorem and its functional form, i.e.
convergence to fractional Brownian motion, for the case when the innovations
are in the domain of attraction of the normal distribution and the constants
satisfy (\ref{dcoeff}). The normalizer in this theorem depends on the slowly
varying function $l(x)$ that is in general unknown. To make our results easily
applicable we also study the central limit theorem in its self-normalized form.

The self-normalized CLT\ for sums of independent identically distributed
random variables was treated in the paper by Gin\'{e} et al (1997). The case
of self-normalized sums in the domain of attraction of other stable laws was
considered by Chistyakov and G\"{o}tze (2004). A systematic treatment of
self-normalized limit theory under independence assumption is given in the
book by de la Pe\~{n}a et al (2009). The self-normalized version of the
functional central limit theorem for this case, was treated in Cs\"{o}rg\H{o}
et al (2003). Kulik (2006) studied the self-normalized functional CLT\ when
$\sum_{i\geq0}|a_{i}|<\infty.$ We shall consider the long memory case when
coefficients satisfy (\ref{dcoeff}).

Our paper is organized in the following way: Section 2 contains the
definitions and the results; the proofs are given in section 3. For
convenience, in the Appendix, we give some auxiliary results and we also
mention some known facts needed for the proofs.

In this paper we shall use the following notations: a double indexed sequence
with indexes $n$ and $i$ will be denoted by $a_{ni}$ when no confusion is
possible, and sometimes by $a_{n,i};$ we use the notation $a_{n}\mathbb{\sim
}b_{n}$ instead of $a_{n}/b_{n}\rightarrow1$; for positive sequences, the
notation $a_{n}\ll b_{n}$ replaces Vinogradov symbol $O$ and it means that
$a_{n}/b_{n}$ is bounded; $a_{n}=o(b_{n})$ stays for $a_{n}/b_{n}%
\rightarrow0;$ $[x]$ denotes the integer part of $x$; the notation
$\Rightarrow$ is used for weak convergence, and $\overset{P}{\rightarrow}$
denotes convergence in probability. By $var(X)$ we denote the variance of the
random variable $X$ and by $cov(X,Y)$ the covariance of $X$ and $Y$. The weak
convergence to a constant means convergence in probability. We denote by
$D[0,1]$ the space of all functions on $[0,1]$ which have left-hand limits and
are continuous from the right. $N(0,1)$ denotes a standard normal random variable.

\section{Results}

To introduce our results we define a normalizing sequence in the following
way. Recall (\ref{DA}) and (\ref{dcoeff}). Let $b=\inf\left\{  x\geq
1:l(x)>0\right\}  $, define
\begin{equation}
\eta_{j}=\inf\left\{  s:s\geq b+1,\frac{l(s)}{s^{2}}\leq\frac{1}{j}\right\}
,\;\;\;j=1,2,\cdots\label{defeta}%
\end{equation}
and set
\begin{equation}
B_{n}^{2}:=c_{\alpha}l_{n}n^{3-2\alpha}L^{2}(n)\text{ with }l_{n}=l(\eta_{n})
\label{defBn}%
\end{equation}
where
\begin{equation}
c_{\alpha}=\{\int_{0}^{\infty}[x^{1-\alpha}-\max(x-1,0)^{1-\alpha}%
]^{2}dx\}/(1-\alpha)^{2}\text{ .} \label{defcalpha}%
\end{equation}

\begin{theorem}
\label{clt} Define $\{X_{n};n\geq1\}$ by (\ref{X}) and the random element
$W_{n}(t)=S_{[nt]}/B_{n}$ on the space $D[0,1]$. Assume conditions
(\ref{centered}), (\ref{DA}) and (\ref{dcoeff}) are satisfied. Then,
$W_{n}(t)$ converges weakly on the space $D[0,1]$ endowed with Skorohod
topology to the fractional Brownian motion $W_{H}$ with Hurst index
$H=3/2-\alpha$. \newline In particular, for $t=1,$ we have that $S_{n}/B_{n}$
converges in distribution to a standard normal variable.
\end{theorem}

\begin{remark}
In a forthcoming paper the authors treat the central limit theorem for the
situation when $B_{n}^{2}$ is not necessarily regularly varying. However, for
that situation the convergence to the fractional Brownian motion might fail.
As a matter of fact, in the context of Theorem \ref{clt} a necessary condition
for the convergence to the fractional Brownian motion $W_{H}$ with Hurst index
$H=\beta$ is the representation $B_{n}^{2}=n^{2\beta}h(x)$ for a function
$h(x)$ that is slowly varying at infinity (see Lamperti, 1962).
\end{remark}

For successfully applying this theorem we have to know $l_{n}$ that depends on
the distribution of $\varepsilon.$ This can be avoided by constructing a
selfnormalizer. Denote $\sum_{i=0}^{\infty}a_{i}^{2}=A^{2}.$ Our result is:

\begin{theorem}
\label{selfCLT}Under the same conditions as in Theorem \ref{clt} we have
\begin{equation}
\frac{1}{nl_{n}}\sum_{i=1}^{n}X_{i}^{2}\overset{P}{\rightarrow}A^{2}
\label{LLN}%
\end{equation}
and therefore%
\[
\frac{S_{[nt]}}{na_{n}\sqrt{\sum_{i=1}^{n}X_{i}^{2}}}\Rightarrow\frac
{\sqrt{c_{\alpha}}}{A}W_{H}(t)\text{ .}%
\]
In particular
\[
\frac{S_{n}}{na_{n}\sqrt{\sum_{i=1}^{n}X_{i}^{2}}}\Rightarrow N(0,\frac
{c_{\alpha}}{A^{2}})\text{ .}%
\]

\end{theorem}

\section{Application to unit root testing}

Invariance principles play an important role in characterizing the limit
distribution of various statistics arising from the inference in economic time series.

Let us consider a stochastic process generated according to%

\[
Y_{n}=\rho Y_{n-1}+X_{n}\text{ for }n\geq1
\]
where $Y_{0}=0$ and $(X_{n})_{n\geq1}$ is a stationary sequence and $\rho$ is
a constant. Denote the ordinary least squares (OLS) estimator of $\rho$ by
\[
\hat{\rho}_{n}=\sum_{k=1}^{n}Y_{k}Y_{k-1}/\sum_{k=1}^{n}Y_{k-1}^{2}\text{ .}%
\]
To test $\rho=1$ against $\rho<1$, a key step is to derive the limit
distribution of the well-known Dickey--Fuller (DF) test statistic (Dickey and
Fuller 1979, 1981):%

\[
\hat{\rho}_{n}-1=\sum_{k=1}^{n}Y_{k-1}(Y_{k}-Y_{k-1})/\sum_{k=1}^{n}%
Y_{k-1}^{2}\text{ .}%
\]
As shown by Phillips (1987), under the null hypothesis $\rho=1$, the
asymptotic properties of the DF test statistic rely heavily on the invariance
principles. This problem was widely studied under various assumptions on the
sequence $X_{n}.$ Among them Sowell (1990) and Wu (2006) considered the unit
root testing problem for long-memory processes. By combining our Theorems
\ref{clt} and \ref{selfCLT} with arguments similar to Phillips (1987), we can
formulate the following result obtained for variables that do not necessarily
have finite second moment.

\begin{proposition}
Assume that $(X_{n})_{n\geq1}$ is as in Theorem \ref{clt}. Then the following
results hold%
\[
(a)\text{ \ \ \ \ \ \ \ \ \ \ \ \ \ \ \ \ \ \ \ }\frac{\sum_{k=1}^{n}%
Y_{k-1}^{2}}{n^{3}a_{n}^{2}\sum_{i=1}^{n}X_{i}^{2}}\Rightarrow\text{\ }%
\frac{c_{\alpha}}{A^{2}}\int_{0}^{1}W_{H}^{2}(t)dt\text{ .\ }%
\]%
\[
(b)\text{ \ \ \ \ \ \ \ \ \ \ \ \ \ \ \ \ \ \ }\frac{\sum_{k=1}^{n}%
Y_{k-1}(Y_{k}-Y_{k-1})}{n^{2}a_{n}^{2}\sum_{i=1}^{n}X_{i}^{2}}\Rightarrow
\frac{c_{\alpha}W_{H}^{2}(1)}{2A^{2}}\text{ .}%
\]%
\[
(c)\text{ \ \ \ \ \ \ \ \ \ \ \ \ \ \ \ \ \ \ }n(\hat{\rho}_{n}-1)\Rightarrow
\frac{W_{H}^{2}(1)/2}{\int_{0}^{1}W_{H}^{2}(t)dt}\text{ }.\text{
\ \ \ \ \ \ \ \ \ \ \ \ }%
\]

\end{proposition}

The proof of this proposition requires only to make obvious changes in the
proofs of (A1) and (A2) on page 296 in Phillips (1987), and it is left to the reader.

\section{Proofs}

\subsection{Proof of theorem \ref{clt}}

In order to prove the central limit theorem in its functional form,\textbf{
}i.e. the weak convergence of $S_{[nt]}/B_{n}$ on the space $D[0,1]$ to the
fractional Brownian motion $W_{H}$ with Hurst index $H=3/2-\alpha$, we shall
first reduce the problem to truncated random variables. For the truncated
process we establish tightness on $D[0,1]$ and the convergence of finite
dimensional distributions.

Without the loss of generality, in the rest of the paper, we assume for
convenience $a_{0}=0$ in definition (\ref{X}).

We shall divide the proof in several steps:

\textbf{Step 1. Existence.}

To show that $X_{1}$ is well defined we use stationarity and Lemma
\ref{csorgo} from the Appendix. First of all we have
\[
\sum_{i=1}^{\infty}P(|a_{i}\varepsilon_{1-i}|>1)=\sum_{i=1}^{\infty
}P(|\varepsilon|>|a_{i}|^{-1})=\sum_{i=1}^{\infty}a_{i}^{2}o(l(|a_{i}%
|^{-1}))\text{ .}%
\]
Then, by taking into account that (\ref{centered}) implies $E\varepsilon
I(|\varepsilon|\leq|a_{i}|^{-1})=-E\varepsilon I(|\varepsilon|>|a_{i}%
|^{-1}),$
\[
\sum_{i=1}^{\infty}|Ea_{i}\varepsilon_{1-i}I(|a_{i}\varepsilon_{1-i}%
|\leq1)|\leq\sum_{i=1}^{\infty}|a_{i}|E|\varepsilon|I(|\varepsilon
|>|a_{i}|^{-1})=\sum_{i=1}^{\infty}a_{i}^{2}o(l(|a_{i}|^{-1}))\text{ }%
\]
and%
\[
\sum_{i=1}^{\infty}Ea_{i}^{2}\varepsilon_{1-i}^{2}I(|a_{i}\varepsilon
_{1-i}|\leq1)=\sum_{i=1}^{\infty}a_{i}^{2}E\varepsilon^{2}I(|\varepsilon
|\leq|a_{i}|^{-1})=\sum_{i=1}^{\infty}a_{i}^{2}l(|a_{i}|^{-1})\text{ .}%
\]
Notice that%
\[
\sum_{i=1}^{\infty}a_{i}^{2}l(|a_{i}|^{-1})=\sum_{i=1}^{\infty}i^{-2\alpha
}L^{2}(i)l(i^{2\alpha}L^{-2}(i))<\infty\text{ ,}%
\]
since $1/2<\alpha<1$ and $L^{2}(i)l(i^{2\alpha}L^{-2}(i))$ is a slowly varying
function at $\infty$. The existence in the almost sure sense follows by
combining these arguments with the three series theorem.

\textbf{Step 2}. \textbf{Truncation.}

For the case when$~E\varepsilon^{2}=\infty$, which is relevant to our paper,
the truncation is necessary. The challenge is to find a suitable level of
truncation. For any integer $1\leq k\leq n$ define
\begin{equation}
X_{nk}^{\prime}=\sum_{i=1}^{\infty}a_{i}\varepsilon_{k-i}I(|\varepsilon
_{k-i}|\leq\eta_{n-k+i})\text{ and }S_{n}^{\prime}=\sum_{k=1}^{n}%
X_{nk}^{\prime}\text{ .} \label{defprime}%
\end{equation}
This definition has the advantage that $S_{n}^{\prime}$ can be expressed as a
simple sum of a linear process of an array of independent variables. For every
$m\geq1$ we denote%
\begin{equation}
b_{m}=a_{1}+\ldots+a_{m}\text{ }, \label{defbm}%
\end{equation}
and then we introduce the coefficients
\begin{align}
b_{nj}  &  =b_{j}=a_{1}+\ldots+a_{j}\text{ for }j<n\label{defbn}\\
b_{nj}  &  =b_{j}-b_{j-n}=a_{j-n+1}+\ldots+a_{j}\,\text{\ for }j\geq n\text{
.}\nonumber
\end{align}
With this notation and recalling definition (\ref{defeta}), by changing the
order of summation,
\begin{equation}
S_{n}^{\prime}=\sum_{i\geq1}b_{ni}\varepsilon_{n-i}I(|\varepsilon_{n-i}%
|\leq\eta_{i})\text{ .} \label{repS}%
\end{equation}
We shall reduce next, the study of limiting distribution of $S_{n}/B_{n}$ to
the sequence $S_{n}^{\prime}/B_{n}.$ It is enough to show that
\begin{equation}
\frac{1}{B_{n}}E|S_{n}-S_{n}^{^{\prime}}|\rightarrow0\text{ .} \label{L1}%
\end{equation}
To see this we use the fact that by Lemma \ref{csorgo} stated in the Appendix
\[
E|\varepsilon|I(|\varepsilon|>\eta_{i})=o(\eta_{i}^{-1}l_{i})\text{ .}%
\]
We also know that
\begin{equation}
\eta_{n}^{2}\sim nl_{n}\text{ .} \label{ordereta}%
\end{equation}
(see for instance relation 13 in Cs\"{o}rg\H{o} et al, 2003).\ Then, by the
triangle inequality and relation (\ref{boundb}) of Lemma \ref{coeff} from the
Appendix applied with $p=1$, we obtain
\begin{gather}
E|S_{n}-S_{n}^{^{\prime}}|\leq\sum_{i\geq1}|b_{ni}|E|\varepsilon
|I(|\varepsilon|>\eta_{i})=\sum_{i\geq1}|b_{ni}|o(\eta_{i}^{-1}l_{i}%
)\label{negl2}\\
=\sum_{i\geq1}|b_{ni}|o(i^{-1/2}l_{i}^{1/2})=o(n^{3/2-\alpha}l_{n}%
^{1/2}L(n))=o(B_{n})\nonumber
\end{gather}
and so (\ref{L1}) is established.

\textbf{Step} \textbf{3.} \textbf{Central limit theorem.}

To make the proof more transparent we shall present first the central limit
theorem for $S_{n}/B_{n}$. By the Step 2 it is enough to find the limiting
distribution of $S_{n}^{^{\prime}}/B_{n}$. We start by noticing that by
(\ref{negl2}) and the fact that the variables are centered we have
\begin{equation}
|ES_{n}^{^{\prime}}|=|E(S_{n}-S_{n}^{^{\prime}})|=o(B_{n})\text{ .}
\label{negl5}%
\end{equation}
One of the consequences of this observation is that $S_{n}^{^{\prime}}/B_{n}$
has the same limiting distribution as $(S_{n}^{^{\prime}}-ES_{n}^{\prime
})/B_{n}$. Furthermore,
\[
var(\frac{S_{n}^{^{\prime}}}{B_{n}})=\frac{1}{B_{n}^{2}}\sum_{i\geq1}%
b_{ni}^{2}l_{i}-\frac{1}{B_{n}^{2}}(ES_{n}^{^{\prime}})^{2}\rightarrow1
\]
by relation (\ref{VAR}) in Lemma \ref{coeff} and (\ref{negl5}).

Moreover, by the point 1 in Lemma \ref{coeff} for $k_{n}=n^{4/(2\alpha-1)}$%
\begin{align}
var(\sum_{i\geq k_{n}}b_{ni}\varepsilon_{n-i}I(|\varepsilon_{n-i}|  &
\leq\eta_{i}))\ll\sum_{i\geq k_{n}}n^{2}(i-n)^{-2\alpha}L^{2}(i)l_{i}%
\label{n}\\
&  =o(1)\text{ as }n\rightarrow\infty\text{ .}\nonumber
\end{align}
Then, by Theorem 4.1 in Billingsley (1968), for proving the central limit
theorem it is enough to verify Lyapunov's condition for $B_{n}^{-1}(\bar
{S}_{n}^{\prime}-E\bar{S}_{n}^{\prime})$ where%
\[
\bar{S}_{n}^{\prime}=\sum_{i=1}^{k_{n}}b_{ni}\varepsilon_{n-i}I(|\varepsilon
_{n-i}|\leq\eta_{i})\text{ .}%
\]
Clearly, by (\ref{n}), $var(\bar{S}_{n}^{\prime}/B_{n})\rightarrow1$. In the
estimate below we use the point 4 of Lemma \ref{csorgo} along with
(\ref{ordereta}), followed by relation (\ref{boundb}) of Lemma \ref{coeff}
applied with $p=3$ and the fact that $B_{n}\rightarrow\infty$ to get:
\begin{gather}
\sum_{j=1}^{k_{n}}|b_{nj}|^{3}E|\varepsilon^{\prime}-E\varepsilon^{\prime
}|^{3}\leq8\sum_{j=1}^{k_{n}}|b_{nj}|^{3}E|\varepsilon|^{3}I(|\varepsilon
|\leq\eta_{j})\label{Leap}\\
=\sum_{j=1}^{k_{n}}|b_{nj}|^{3}\eta_{j}o(l_{j})\leq\sum_{j=1}^{\infty}%
|b_{nj}|^{3}j^{1/2}o(l_{j}^{3/2})=o(n^{3(3/2-\alpha)}l_{n}^{3/2}%
L^{3}(n))=o(B_{n}^{3})\text{ .}\nonumber
\end{gather}
By Lyapunov's central limit theorem and the above considerations, $S_{n}%
/B_{n}$ converges to $N(0,1)$ in distribution.

\textbf{Step 4. Preliminary considerations for the convergence to fractional
Brownian motion.}

For $n\geq1$ fixed we implement the same level of truncation as before and
construct $\{X_{nj}^{\prime};1\leq j\leq n\}$ by definition (\ref{defprime}).
Then we introduce the processes
\[
W_{n}^{\prime}(t)=\frac{1}{B_{n}}\sum_{j=1}^{[nt]}X_{nj}^{\prime}\text{ and
}W_{n}^{^{\prime\prime}}(t)=W_{n}(t)-W_{n}^{\prime}(t)\text{ .}%
\]
We shall show first that $W_{n}^{^{\prime\prime}}(t)$ is negligible for the
weak convergence on $D[0,1]$ and then, in the next steps, that $W_{n}^{\prime
}(t)$ is weakly convergent to the fractional Brownian motion.

In order to explain this step, it is convenient to express the process in an
expanded form. By using notation (\ref{defbm})
\[
W_{n}^{^{\prime\prime}}(t)=\frac{1}{B_{n}}\sum_{i=0}^{[nt]-1}b_{[nt]-i}%
\varepsilon_{i}I(|\varepsilon_{i}|>\eta_{n-i})+\frac{1}{B_{n}}\sum_{i\geq
1}(b_{[nt]+i}-b_{i})\varepsilon_{-i}I(|\varepsilon_{-i}|>\eta_{n+i})\text{ .}%
\]
We notice that by the triangle inequality,
\begin{gather*}
E(\sup_{0\leq t\leq1}|W_{n}^{^{\prime\prime}}(t)|)\leq\frac{1}{B_{n}}%
E(\sup_{0\leq t\leq1}|\sum_{i=0}^{[nt]-1}b_{[nt]-i}\varepsilon_{i}%
I(|\varepsilon_{i}|>\eta_{n-i})|)\\
+\frac{1}{B_{n}}E(\sup_{0\leq t\leq1}|\sum_{i\geq1}(b_{[nt]+i}-b_{i}%
)\varepsilon_{-i}I(|\varepsilon_{-i}|>\eta_{n+i})|)\text{ .}%
\end{gather*}
Then, by monotonicity and using the notation (\ref{defbn})%
\begin{gather}
E(\sup_{0\leq t\leq1}|W_{n}^{^{\prime\prime}}(t)|)\leq\frac{1}{B_{n}}%
\sum_{i=0}^{n-1}|b_{n-i}|E|\varepsilon|I(|\varepsilon|>\eta_{n-i}%
)\label{negl3}\\
+\frac{1}{B_{n}}\sum_{i\geq1}|b_{n+i}-b_{i}|E|\varepsilon|I(|\varepsilon
|>\eta_{n+i})=\frac{1}{B_{n}}\sum_{i\geq1}|b_{ni}|E|\varepsilon|I(|\varepsilon
|>\eta_{i})\text{ .}\nonumber
\end{gather}
which is exactly the quantity shown to converge to $0$ in (\ref{negl2}). By
Theorem 4.1 in Billingsley (1968), it is enough to study the limiting behavior
of $W_{n}^{\prime}(t)$.

\textbf{Step 5. Tightness.}

As before, we reduce the problem to studying the same problem for
$W_{n}^{\prime}(t)-EW_{n}^{\prime}(t).$ This is easy to see, since, by the
fact the variables are centered and by (\ref{negl3}) we clearly obtain
\begin{equation}
\sup_{0\leq t\leq1}|EW_{n}^{\prime}(t)|=\sup_{0\leq t\leq1}|EW_{n}%
^{^{\prime\prime}}(t)|\leq E(\sup_{0\leq t\leq1}|W_{n}^{^{\prime\prime}%
}(t)|)\rightarrow0\text{ .} \label{negl4}%
\end{equation}
In order to show that $W_{n}^{\prime}(t)-EW_{n}^{\prime}(t)$ is tight in
$D[0,1]$ we shall verify the conditions from Lemma \ref{Ltight}, in Appendix,
for the triangular array $B_{n}^{-1}(X_{nk}^{\prime}-EX_{nk}^{\prime})$,
$1\leq k\leq n.$ This will be achieved in the following two lemmas.

By the properties of slowly varying functions (see Seneta 1976 and also Lemma
\ref{Karamata} in Appendix) we construct first an integer $N_{0}$ and positive
constants $K_{i}$ such that for all $m>N_{0}$ we have simultaneously
\begin{equation}
\max_{1\leq j\leq m}b_{j}^{2}\leq K_{1}m^{2-2\alpha}L^{2}(m)\text{ ,}
\label{1}%
\end{equation}%
\begin{equation}
l_{2m}\leq K_{2}l_{m}\text{ ,} \label{2}%
\end{equation}%
\begin{equation}
\sup_{k>2m}\frac{(b_{k}-b_{k-m})^{2}}{k^{-2\alpha}L^{2}(k)}\leq K_{3}%
m^{2}\text{ ,} \label{3}%
\end{equation}%
\begin{equation}
\sum_{j\geq m}j^{-2\alpha}L^{2}(j)l_{j}\leq K_{4}m^{1-2\alpha}L^{2}%
(m)l_{m}\text{ .} \label{4}%
\end{equation}
and%
\begin{equation}
\sum_{j\geq m}j^{-2\alpha}L^{2}(j)\leq K_{4}m^{1-2\alpha}L^{2}(m)\text{ .}
\label{5}%
\end{equation}

This is possible by Lemma \ref{coeff} \ and Lemma \ref{Karamata}.

\begin{lemma}
There is a constant $K$ and an integers $N_{0}$ such that for any two integers
$p$ and $q$ with $1\leq p<q\leq n$ with $q-p\geq N_{0}$ and any $n\geq
N_{0}\ $
\begin{equation}
\frac{1}{B_{n}^{2}}var(\sum_{i=p+1}^{q}X_{ni}^{\prime})\leq K(\frac{q}%
{n}-\frac{p}{n})^{2-\alpha}\text{ .} \label{tight}%
\end{equation}

\end{lemma}

\textbf{Proof.} We shall use $N_{0}$ that was already constructed above. We
start from the decomposition
\begin{gather*}
\sum_{i=p+1}^{q}X_{ni}^{\prime}=\sum_{i=p}^{q-1}b_{q-i}\varepsilon
_{i}I(|\varepsilon_{i}|\leq\eta_{n-i})+\sum_{i=2p-q}^{p-1}(b_{q-i}%
-b_{p-i})\varepsilon_{i}I(|\varepsilon_{i}|\leq\eta_{n-i})\\
+\sum_{i\geq q-2p+1}(b_{q+i}-b_{p+i})\varepsilon_{-i}I(|\varepsilon_{-i}%
|\leq\eta_{n+i})=I+II+III\text{ .}%
\end{gather*}
We shall estimate the variance of each term separately.

Using the fact that $l_{n}$ is increasing and (\ref{1}) we obtain
\begin{align*}
var(I)  &  \leq\sum_{i=p}^{q-1}b_{q-i}^{2}l_{n-i}=\sum_{j=1}^{q-p}b_{j}%
^{2}l_{n-q+j}\leq l_{n}(q-p)\max_{1\leq j\leq q-p}b_{j}^{2}\\
&  \leq K_{1}(q-p)^{3-2\alpha}L^{2}(q-p)l_{n}\text{ }.
\end{align*}
Then, by taking into account that $l_{n}$ is increasing, (\ref{1}) and
(\ref{2}) we have
\begin{gather*}
var(II)\leq\sum_{i=2p-q}^{p-1}(b_{q-i}-b_{p-i})^{2}l_{n-i}\leq l_{2n}%
2(q-p)\max_{1\leq j\leq2(q-p)}b_{i}^{2}\\
\leq K_{1}K_{2}(q-p)^{3-2\alpha}L^{2}(q-p)l_{n}\text{ .}%
\end{gather*}
To estimate the variance of the last term, we use first (\ref{3}) to obtain
\begin{gather*}
var(III)=\sum_{i\geq q-2p+1}(b_{i+q}-b_{i+p})^{2}l_{n+i}\leq\sum
_{j\geq2(q-p)+1}(b_{j}-b_{j-(q-p)})^{2}l_{n+j-q}\\
\leq K_{3}(q-p)^{2}\sum_{j\geq2(q-p)+1}j^{-2\alpha}L^{2}(j)l_{n+j-q}\text{ }.
\end{gather*}
Now, by the monotonicity of $l_{n}$, because $l_{n+j-q}\leq l_{2n}$ for $j\leq
n$ and $l_{n+j-q}\leq l_{2j}$ for $j>n$ by (\ref{2}), (\ref{4}) and (\ref{5})
\begin{align*}
\sum_{j\geq2(q-p)+1}j^{-2\alpha}L^{2}(j)l_{n+j-q}  &  \leq K_{2}%
K_{5}(q-p)^{-2\alpha+1}L^{2}(q-p)l_{n}\\
&  +K_{2}K_{4}(q-p)^{-2\alpha+1}L^{2}(q-p)l_{q-p}%
\end{align*}
So, for $K_{6}=K_{2}K_{3}(K_{4}+K_{5})$
\[
var(III)\leq K_{6}(q-p)^{3-2\alpha}L^{2}(q-p)l_{n}\text{ .}%
\]
Overall we have so far for a certain constant $K_{7}$ that does not depend on
$p$ or $q$,
\begin{equation}
var(\sum_{i=p+1}^{q}X_{ni}^{\prime})\leq K_{7}(q-p)^{3-2\alpha}L^{2}%
(q-p)l_{n}\text{ .} \label{Ineqvar}%
\end{equation}
By simple algebra, because $1\leq p<q\leq n$ we derive
\[
var(\sum_{i=p+1}^{q}X_{ni}^{\prime})\leq K_{7}(q-p)^{2-\alpha}l_{n}%
n^{1-\alpha}L^{2}(n)\max_{1\leq k\leq n}\frac{k^{1-\alpha}}{n^{1-\alpha}}%
\frac{L^{2}(k)}{L^{2}(n)}\text{ }.
\]
Finally, by the point 5 of Lemma \ref{Karamata},
\[
var(\sum_{i=p+1}^{q}X_{ni}^{\prime})\leq K_{8}(\frac{q}{n}-\frac{p}%
{n})^{2-\alpha}l_{n}n^{3-2\alpha}L^{2}(n)\text{ .}%
\]
Therefore, (\ref{tight}) is established by taking into account (\ref{defBn}).
$\lozenge$

\begin{lemma}
Condition (\ref{condtight1}) is satisfied, namely:
\[
\lim_{n\rightarrow\infty}P(\max_{1\leq k\leq n}|X_{nk}^{\prime}-EX_{nk}%
^{\prime}|\geq\varepsilon B_{n})=0\text{ .}%
\]

\end{lemma}

\textbf{Proof.} We start from%

\[
P(\max_{1<k\leq n}|X_{nk}^{\prime}-EX_{nk}^{\prime}|\geq\varepsilon B_{n}%
)\leq\frac{1}{\varepsilon^{4}B_{n}^{4}}\sum_{k=1}^{n}E|X_{nk}^{\prime}%
-EX_{nk}^{\prime}|^{4}\text{ .}%
\]
We use now Rosenthal inequality (Theorem 1.5.13 in de la Pe\~{n}a and Gin\'{e}
1999), which can be easily extended to an infinite sum of independent random
variables, by truncating the sum and passing to the limit. So, there is a
constant $C,$ such that
\[
E|X_{nk}^{\prime}-EX_{nk}^{\prime}|^{4}\leq C\sum_{i=1}^{\infty}a_{i}%
^{4}E\varepsilon^{4}I(|\varepsilon|\leq\eta_{n-k+i})+C(\sum_{i=1}^{\infty
}a_{i}^{2}l_{n-k+i})^{2}=I_{k}+II_{k}\text{ .}%
\]
By the point 4 of Lemma \ref{csorgo} and (\ref{ordereta}) it follows that%
\begin{align*}
a_{i}^{4}E\varepsilon^{4}I(|\varepsilon|  &  \leq\eta_{n-k+i})\ll a_{i}%
^{4}(\eta_{n-k+i}^{2})l_{n-k+i}\\
&  \ll i^{-4\alpha}L^{4}(i)(n-k+i)l_{n-k+i}^{2}\text{ .}%
\end{align*}
So%
\[
\sum_{k=1}^{n}I_{k}\leq\sum_{i=1}^{\infty}i^{-4\alpha}L^{4}(i)\sum_{k=i}%
^{n+i}kl_{k}^{2}\ll n^{2}l_{n}^{2}\text{ .}%
\]
Then, by simple computations involving the partition of sum in two parts, one
up to $2n$ and the rest, and then using the properties of regularly functions
and the fact that $2\alpha>1$ we obtain
\[
\sum_{k=1}^{n}II_{k}\leq n(\sum_{i=1}^{\infty}i^{-2\alpha}L^{2}(i)l_{n+i}%
)^{2}\leq nl_{n}^{2}\text{ .}%
\]
Finally by (\ref{defBn}) we notice that
\[
\frac{n^{2}l_{n}^{2}}{B_{n}^{4}}\rightarrow0\text{ .}%
\]
$\lozenge$

\textbf{Step 6. Convergence of finite dimensional distributions.}

Let $0\leq t_{1}<t_{2}<\dots<t_{m}\leq1$. We shall show next that the vector
$(W_{n}^{\prime}(t_{j});1\leq j\leq m)$ converges in distribution to the
finite dimensional distributions of a fractional Brownian motion with Hurst
index $3/2-2\alpha$, i.e. of a Gaussian process with covariance structure
$\frac{1}{2}(t^{3-2\alpha}+s^{3-2\alpha}-(t-s)^{3-2\alpha})$ for $s<t$.

By the Cram\'{e}r-Wold device and taking into account (\ref{negl4}) we have to
study the limiting distribution of $\sum_{j=2}^{m}\lambda_{j}(W_{n}^{\prime
}(t_{j})-EW_{n}^{\prime}(t_{j-1}))$, which we express as a weighted sum of
independent random variables. By elementary computations involving similar
arguments used in the proof of step 3, and taking into account (\ref{negl2})
and (\ref{Leap}), we notice that Lyapunov's condition is satisfied and then,
the limiting distribution is normal with the covariance structure that will be
specified next. We compute now the covariance of $W_{n}^{\prime}(s)$ and
$W_{n}^{\prime}(t)$ for $s\leq t$. By simple algebra
\[
cov(W_{n}^{\prime}(t),W_{n}^{\prime}(s))=\frac{1}{2}(var(W_{n}^{\prime
}(t))+var(W_{n}^{\prime}(s))-var(W_{n}^{\prime}(t)-W_{n}^{\prime}(s)))\text{
.}%
\]
We analyze now the variance of $W_{n}^{\prime}(t).$ For each $t$ fixed,
$0\leq$ $t\leq1$
\begin{gather*}
var(W_{n}^{\prime}(t))=\frac{1}{B_{n}^{2}}\sum_{i=0}^{[nt]-1}b_{[nt]-i}%
^{2}(E\varepsilon_{0}^{2}I(|\varepsilon_{0}|\leq\eta_{n-i})-E^{2}%
\varepsilon_{0}I(|\varepsilon_{0}|\leq\eta_{n-i}))\\
+\frac{1}{B_{n}^{2}}\sum_{i\geq1}(b_{[nt]+i}-b_{i})^{2}(E\varepsilon_{0}%
^{2}I(|\varepsilon_{0}|\leq\eta_{n+i})-E^{2}\varepsilon_{0}I(|\varepsilon
_{0}|\leq\eta_{n+i}))\text{ .}%
\end{gather*}
Taking into account $E\varepsilon_{0}I(|\varepsilon_{0}|\leq\eta
_{n-i})=-E\varepsilon_{0}I(|\varepsilon_{0}|>\eta_{n-i}),$ by Lemma
\ref{csorgo} and Lemma \ref{coeff}, after some computations, we obtain
\[
var(W_{n}^{\prime}(t))\sim\frac{1}{B_{n}^{2}}\sum_{i=0}^{[nt]-1}b_{[nt]-i}%
^{2}l_{n-i}+\frac{1}{B_{n}^{2}}\sum_{i\geq1}(b_{[nt]+i}-b_{i})^{2}%
l_{n+i}\text{ .}%
\]
With a similar proof as of relation (\ref{sumb}) of Lemma \ref{coeff}, for
every $0\leq t\leq1$%
\[
var(W_{n}^{\prime}(t))\rightarrow t^{3-2\alpha}%
\]
and for every $0\leq s<t\leq1$
\begin{equation}
var(W_{n}^{\prime}(t)-W_{n}^{\prime}(s))\rightarrow(t-s)^{3-2\alpha}\text{
.}\label{st}%
\end{equation}
Then%
\[
cov(W_{n}^{\prime}(t),W_{n}^{\prime}(s))\rightarrow\frac{1}{2}(t^{3-2\alpha
}+s^{3-2\alpha}-(t-s)^{3-2\alpha})\,\text{,}%
\]
that is the desired covariance structure. $\lozenge$

\subsection{Proof of Theorem \textbf{ \ref{selfCLT}}}

We notice that it is enough to prove only the convergence in (\ref{LLN}). Then
(\ref{dcoeff}), (\ref{defBn}) and (\ref{LLN}) imply
\[
B_{n}^{2}\sim c_{\alpha}n^{2}a_{n}^{2}(\sum_{j=1}^{n}X_{j}^{2})/A^{2}\text{
},
\]
which we combine with Theorem \ref{clt}, via Slutsky's theorem, to obtain the
self-normalized part of the theorem. The proof of (\ref{LLN}) will be
decomposed in several steps.

\textbf{Step 1. Truncation.}

Denote $D_{n}^{2}=A^{2}nl_{n}$. Recall the definition (\ref{defprime}) and set
$X_{nj}^{^{\prime\prime}}=X_{j}-X_{nj}^{^{\prime}}$. In order to prove
(\ref{LLN}) it is enough to establish
\begin{equation}
F_{n}=\sum_{j=1}^{n}(X_{nj}^{^{\prime\prime}})^{2}/D_{n}^{2}\overset
{P}{\rightarrow}0 \label{NEG}%
\end{equation}
and
\begin{equation}
G_{n}=\sum_{j=1}^{n}(X_{nj}^{\prime})^{2}/D_{n}^{2}\overset{P}{\rightarrow
}1\text{ .} \label{LLNtr}%
\end{equation}
To see this we square the decomposition $X_{j}=X_{nj}^{\prime}+X_{nj}%
^{^{\prime\prime}}$; then sum with $j$ from $1$ to $n$, and notice that by the
H\"{o}lder inequality
\[
F_{n}-2(G_{n}F_{n})^{1/2}\leq\frac{1}{D_{n}^{2}}\sum_{j=1}^{n}X_{j}^{2}%
-G_{n}\leq F_{n}+2(G_{n}F_{n})^{1/2}\text{ .}%
\]
\textbf{Step 2. Proof of (\ref{NEG}). }

We start from
\begin{gather*}
\sum_{k=1}^{n}(X_{nk}^{^{\prime\prime}})^{2}=\sum_{k=1}^{n}\sum_{i=1}^{\infty
}a_{i}^{2}\varepsilon_{k-i}^{2}I(|\varepsilon_{k-i}|>\eta_{n-k+i})+\\
2\sum_{k=1}^{n}\sum_{i<j}a_{i}a_{j}\varepsilon_{k-i}I(|\varepsilon_{k-i}%
|>\eta_{n-k+i})\varepsilon_{k-j}I(|\varepsilon_{k-j}|>\eta_{n-k+j})=I+II\text{
.}%
\end{gather*}
(here and below $\sum_{i<j}$ denotes double summation). By independence,
monotonicity, and the point 3 of Lemma \ref{csorgo}, we easily deduce that%
\begin{gather*}
E|II|\leq2\sum_{k=1}^{n}\sum_{i<j}|a_{i}a_{j}|E|\varepsilon_{k-i}%
I(|\varepsilon_{k-i}|>\eta_{n-k+i})\varepsilon_{k-j}I(|\varepsilon_{k-j}%
|>\eta_{n-k+j})|\\
\leq2n\sum_{i<j}|a_{i}a_{j}|E|\varepsilon|I(|\varepsilon|>\eta_{i}%
)E|\varepsilon|I(|\varepsilon|>\eta_{j})=2n\sum_{i<j}|a_{i}a_{j}|o(\eta
_{i}^{-1}l_{i})o(\eta_{j}^{-1}l_{j})\ \text{.}%
\end{gather*}
Then, by (\ref{ordereta}), clearly
\[
E|II|\leq n(\sum_{i\geq1}|a_{i}|i^{-1/2}o(l_{i}^{1/2}))^{2}\text{ .}%
\]
Because $\sum_{i\geq1}|a_{i}|i^{-1/2}<\infty,$ and $l_{n\text{ }}$is
increasing, it is easy to see that%
\[
E|II|=o(nl_{n})=o(D_{n}^{2})\text{ .}%
\]
In order to estimate the contribution of the term $I,$ by changing the order
of summation we express this term\ in the following way%
\[
I=\sum_{j=1}^{n}(\sum_{i=1}^{j}a_{i}^{2})\varepsilon_{n-j}^{2}I(|\varepsilon
_{n-j}|>\eta_{j})+\sum_{j=n+1}^{\infty}(\sum_{i=j-n+1}^{j}a_{i}^{2}%
)\varepsilon_{n-j}^{2}I(|\varepsilon_{n-j}|>\eta_{j})\text{ .}%
\]
We implement now the notation
\begin{equation}
A_{nj}^{2}=A_{j}^{2}=\sum_{i=1}^{j}a_{i}^{2}\text{ when }j\leq n\text{ and
}A_{nj}^{2}=\sum_{i=j-n+1}^{j}a_{i}^{2}\text{ when }j>n\text{ .} \label{defA}%
\end{equation}
and then we express $I$ as
\[
I=\sum_{j=1}^{\infty}A_{nj}^{2}\varepsilon_{n-j}^{2}I(|\varepsilon_{n-j}%
|>\eta_{i})\text{ .}%
\]
Clearly $A_{nj}^{2}$ are uniformly bounded by a constant. In addition, by
relation (\ref{propL}), for $j>2n$, these coefficients have the following
order of magnitude%
\begin{align}
A_{nj}^{2}  &  \ll n^{2}(j-n)^{-2\alpha-1}\max_{j-n\leq k\leq j}L^{2}(k)\leq
n^{2}(j-n)^{-2\alpha-1}\max_{j/2\leq k\leq j}L^{2}(k)\label{orderA}\\
&  \ll n^{2}(j-n)^{-2\alpha-1}\min_{j/2\leq k\leq j}L^{2}(k)\leq
n^{2}(j-n)^{-2\alpha-1}L^{2}(j)\text{ .}\nonumber
\end{align}
Now, we use first the Khinchin's inequality (see Lemma 1.4.13 in de la
Pe\~{n}a and Gin\'{e}, 1999) followed by the triangle inequality and Lemma
\ref{csorgo}, and relation (\ref{ordereta}) to obtain
\[
E\sqrt{I}\ll E|\sum_{j=1}^{\infty}A_{nj}\varepsilon_{n-j}I(|\varepsilon
_{n-j}|>\eta_{j})|=\sum_{j=1}^{\infty}A_{nj}o(\eta_{j}^{-1}l_{j})=\sum
_{j=1}^{\infty}A_{nj}j^{-1/2}o(l_{j}^{1/2})\text{ .}%
\]
We notice that by (\ref{orderA}), the point 4 of Lemma \ref{Karamata}, and the
fact that $\alpha>1/2,$%
\begin{align*}
E\sqrt{I}  &  =o(\sqrt{nl_{n}})+n\sum_{j\geq n}^{\infty}j^{-\alpha-1/2}%
L^{2}(j+n)j^{-1/2}o(l_{j}^{1/2})\\
&  =o(\sqrt{nl_{n}})+O(n^{-\alpha}L^{2}(n)l_{j}^{1/2})=o(\sqrt{nl_{n}})\text{
.}%
\end{align*}
As a consequence, $\sqrt{I/nl_{n}}$ converges in $L_{1}$ to $0$, and so,
$I/D_{n}^{2}$ is convergent to $0$ in probability. By gathering all these
facts we deduce that (\ref{NEG}) holds and the proof is reduced to show that
(\ref{LLNtr}) holds.

\textbf{Step 3. Proof of (\ref{LLNtr}).}

We express the sum of squares as%
\begin{gather*}
\sum_{k=1}^{n}(X_{k}^{\prime})^{2}=\sum_{k=1}^{n}\sum_{i=1}^{\infty}a_{i}%
^{2}\varepsilon_{k-i}^{2}I(|\varepsilon_{k-i}|\leq\eta_{n-k+i})\\
+2\sum_{k=1}^{n}\sum_{1\leq i<j}a_{i}a_{j}\varepsilon_{k-i}I(|\varepsilon
_{k-i}|\leq\eta_{n-k+i})\varepsilon_{k-j}I(|\varepsilon_{k-j}|\leq\eta
_{n-k+j})\text{ .}%
\end{gather*}
We shall show that
\begin{equation}
\frac{1}{D_{n}^{2}}\sum_{k=1}^{n}\sum_{i=1}^{\infty}a_{i}^{2}\varepsilon
_{k-i}^{2}I(|\varepsilon_{k-i}|\leq\eta_{n-k+i})\overset{P}{\rightarrow
}1\text{ } \label{convP}%
\end{equation}
and
\begin{equation}
\frac{1}{D_{n}^{2}}\sum_{k=1}^{n}\sum_{1\leq i<j}a_{i}a_{j}\varepsilon
_{k-i}I(|\varepsilon_{k-i}|\leq\eta_{n-k+i})\varepsilon_{k-j}I(|\varepsilon
_{k-j}|\leq\eta_{n-k+j})\overset{P}{\rightarrow}0\text{ .} \label{negl1}%
\end{equation}
We establish first (\ref{convP}).

By using the notation (\ref{defA}), we have
\[
\sum_{k=1}^{n}\sum_{i=1}^{\infty}a_{i}^{2}\varepsilon_{k-i}^{2}I(|\varepsilon
_{k-i}|\leq\eta_{n-k+i})=\sum_{i=1}^{\infty}A_{ni}^{2}\varepsilon_{n-i}%
^{2}I(|\varepsilon_{n-i}|\leq\eta_{i})\text{ .}%
\]
By independence, part 4 of Lemma \ref{csorgo}, relations (\ref{orderA}) and
(\ref{ordereta}), and taking into account that $\alpha>1/2$ we get%
\begin{align*}
&  Var(\sum_{i=1}^{\infty}A_{ni}^{2}\varepsilon_{n-i}^{2}I(|\varepsilon
_{n-i}|\leq\eta_{i}))\leq\sum_{i=1}^{\infty}A_{ni}^{4}E\varepsilon
^{4}\mathbf{1}(|\varepsilon|\leq\eta_{i})\\
&  =\sum_{i=1}^{2n}\eta_{i}^{2}o(l_{i})+\sum_{i\geq2n}(i-n)^{-4\alpha-2}%
L^{4}(i)n^{4}\eta_{i}^{2}o(l_{i})=o(n^{2}l_{n}^{2})=o(D_{n}^{4})\text{ .}%
\end{align*}
So (\ref{convP}) is reduced to showing that
\[
\frac{1}{D_{n}^{2}}\sum_{i=1}^{\infty}A_{ni}^{2}E(\varepsilon^{2}%
I(|\varepsilon|\leq\eta_{i})=\frac{1}{D_{n}^{2}}\sum_{i=1}^{\infty}A_{ni}%
^{2}l_{i}\rightarrow1\text{ as }n\rightarrow\infty\text{ .}%
\]
We divide the sum in three parts, one from $1$ to $n$, one from $n+1$ to $2n$
and the rest of the series. We easily see that by (\ref{orderA}),%

\[
\sum_{j=2n+1}^{\infty}A_{nj}^{2}l_{j}\ll\sum_{j=2n+1}^{\infty}(j-n)^{-2\alpha
-1}n^{2}L^{2}(j)l_{j}=o(D_{n}^{2})\text{ .}%
\]
Then,
\begin{gather*}
\sum_{j=n+1}^{2n}A_{nj}^{2}l_{j}\ll\sum_{j=n+1}^{2n}(j-n)^{1-2\alpha}l_{j}%
\max_{1\leq i\leq n}L(i)\\
\ll(\sum_{k=1}^{n}k^{1-2\alpha}l_{n+k})\max_{1\leq i\leq n}L(i)=o(D_{n}%
^{2})\text{ .}%
\end{gather*}
Now, by the proof of relation (\ref{rel1}) in Appendix with the only
difference that we replace $a_{i}$ by $a_{i}^{2}$ and so $b_{n}^{2}$ by
$A_{j}^{2}$ we obtain%
\begin{equation}
\sum_{j=1}^{n}A_{j}^{2}l_{j}\sim l_{n}\sum_{j=1}^{n}A_{j}^{2}\text{ .}
\label{relA}%
\end{equation}
Finally, by (\ref{relA}), the definition of $D_{n}^{2}$ and by the Toeplitz
lemma, (\ref{toeplitz}) in Appendix, it follows%
\[
\lim_{n\rightarrow\infty}\frac{1}{D_{n}^{2}}\sum_{j=1}^{n}A_{j}^{2}l_{j}%
=\lim_{n\rightarrow\infty}\frac{l_{n}\sum_{j=1}^{n}A_{j}^{2}}{l_{n}nA^{2}%
}=\lim_{n\rightarrow\infty}\frac{A_{n}^{2}}{A^{2}}=1\text{ .}%
\]
This completes the proof of (\ref{convP}).

We move now to prove (\ref{negl1}). Let $N$ be a fixed positive integer. For
each $1\leq k\leq n$ we divide the sum in two parts:
\begin{align*}
\sum_{j=1}^{N}\sum_{i=1}^{j-1}a_{i}a_{j}\varepsilon_{k-i}I(|\varepsilon
_{k-i}|  &  \leq\eta_{n-k+i})\varepsilon_{k-j}I(|\varepsilon_{k-j}|\leq
\eta_{n-k+j})\\
+\sum_{j>N}\sum_{i=1}^{j-1}a_{i}a_{j}\varepsilon_{k-i}I(|\varepsilon_{k-i}|
&  \leq\eta_{n-k+i})\varepsilon_{k-j}I(|\varepsilon_{k-j}|\leq\eta
_{n-k+j})=I_{k}+II_{k}\text{ .}%
\end{align*}
We estimate the variance of the sum of each term separately.

For estimating $var(\sum_{k=1}^{n}II_{k})$ we apply the H\"{o}lder inequality:%
\[
var(\sum_{k=1}^{n}II_{k})\leq n\sum_{k=1}^{n}var(\sum_{j>N}\sum_{i=1}%
^{j-1}a_{i}a_{j}\varepsilon_{k-i}I(|\varepsilon_{k-i}|\leq\eta_{n-k+i}%
)\varepsilon_{k-j}I(|\varepsilon_{k-j}|\leq\eta_{n-k+j}))\text{ .}%
\]
By independence, a term corresponding to the combination of indexes
$(k-i_{1},k-j_{1},k-i_{2},k-j_{2})$ with $i_{1}<j_{1}$ has a non-null
contribution if and only if $i_{1}=i_{2}$ and $j_{1}=j_{2},$ leading to
\[
var(\sum_{k=1}^{n}II_{k})\leq n^{2}\sum_{j>N}\sum_{i=1}^{j-1}a_{i}^{2}%
a_{j}^{2}l_{n+i}l_{n+j}=(n^{2}l_{n}^{2})o_{N}(1)\text{ ,}%
\]
where we used first the monotonicity of $l_{n}$ and in the last part we used
the fact that (by monotonicity, the definition of slowly varying functions and
our notations) $l_{i}\leq l_{2n}\ll l_{n},$ for $i\leq2n$ and $l_{i+n}\leq
l_{3i/2}\ll l_{i}$ for $i>2n$ along with the convergence of the series
$\sum_{i}a_{i}^{2}l_{i}$.

In order to treat the other term we start from%
\[
var(\sum_{k=1}^{n}I_{k})=var(\sum_{j=1}^{N}\sum_{i=1}^{j-1}a_{i}a_{j}%
\sum_{k=1}^{n}\varepsilon_{k-i}I(|\varepsilon_{k-i}|\leq\eta_{n-k+i}%
)\varepsilon_{k-j}I(|\varepsilon_{k-j}|\leq\eta_{n-k+j}))
\]
and then, because we compute the variance of at most $N^{2}$ sums and because
the coefficients $a_{i}$ are bounded, clearly,%
\[
var(\sum_{k=1}^{n}I_{k})\ll N^{4}\max_{1\leq i<j\leq N}var(\sum_{k=1}%
^{n}\varepsilon_{k-i}I(|\varepsilon_{k-i}|\leq\eta_{n-k+i})\varepsilon
_{k-j}I(|\varepsilon_{k-j}|\leq\eta_{n-k+j}))\text{ .}%
\]
We notice now that
\begin{gather*}
var(\sum_{k=1}^{n}\varepsilon_{k-i}I(|\varepsilon_{k-i}|\leq\eta
_{n-k+i})\varepsilon_{k-j}I(|\varepsilon_{k-j}|\leq\eta_{n-k+j}))\\
\leq\sum_{k=1}^{n}E\varepsilon_{k-i}^{2}I(|\varepsilon_{k-i}|\leq\eta
_{n-k+i})E\varepsilon_{k-j}^{2}I(|\varepsilon_{k-j}|\leq\eta_{n-k+j}))\text{
,}%
\end{gather*}
since by independence and the fact that $i\neq j$ all the other terms are
equal to $0$. The result is%
\[
var(\sum_{k=1}^{n}I_{k})\ll N^{4}\sum_{k=1}^{n}l_{n,n-k+i}l_{n,n-k+j}\ll
N^{4}(nl_{n}^{2})\text{ .}%
\]
Overall%
\begin{gather*}
\frac{1}{D_{n}^{4}}var(\sum_{k=1}^{n}\sum_{1\leq i<j}\varepsilon
_{k-i}I(|\varepsilon_{k-i}|\leq\eta_{n-k+i})\varepsilon_{k-j}I(|\varepsilon
_{k-j}|\leq\eta_{n-k+j}))\leq\\
\frac{2}{D_{n}^{4}}var(\sum_{k=1}^{n}I_{k})+\frac{2}{D_{n}^{4}}var(\sum
_{k=1}^{n}II_{k})=o_{N}(1)+O(N^{4}\frac{1}{n})\text{.}%
\end{gather*}
We conclude that (\ref{negl1}) holds by letting first $n\rightarrow\infty$
followed by $N\rightarrow\infty$. $\lozenge$

\section{Appendix}

We formulate in the first lemma several properties of the slowly varying
function. Their proofs can be found in Seneta (1976).

\begin{lemma}
\label{Karamata} A slowly varying function $l(x)$ defined on $[A,\infty)$ has
the following properties:\newline

\begin{enumerate}
\item There exists $B\geq A$ such that for all $x\geq B,$ $l(x)$ is
representable in the form $l(x)=g(x)\exp(\int_{B}^{x}\frac{a(y)}{y}dy)$, where
$g(x)\rightarrow c_{0}>0$, and $a(x)\rightarrow0$ as $x\rightarrow\infty$. In
addition $a(x)$ is continuous.

\item For $B<c<C<\infty$, $\lim_{x\rightarrow\infty}\frac{l(tx)}{l(x)}=1$
uniformly in $c\leq t\leq C$.

\item For any $\theta>-1$, $\int_{B}^{x}y^{\theta}l(y)dy\mathbb{\sim}%
\frac{x^{\theta+1}l(x)}{\theta+1}$ as $x\rightarrow\infty$.

\item For any $\theta<-1$, $\int_{x}^{\infty}y^{\theta}l(y)dy\mathbb{\sim
}\frac{x^{\theta+1}l(x)}{-\theta-1}$ as $x\rightarrow\infty$.

\item For any $\eta>0$, $\sup_{t\geq x}(t^{\eta}l(t))\mathbb{\sim}x^{\eta
}l(x)$ as $x\rightarrow\infty$. Moreover $\sup_{t\geq x}(t^{\eta}%
l(t))=x^{\eta}\bar{l}(x)$ where $\bar{l}(x)$ is slowly varying and $\bar
{l}(x)\mathbb{\sim}l(x).$
\end{enumerate}
\end{lemma}

The following lemma contains some equivalent formulation for variables in the
domains of attraction of normal law (\ref{DA}). It is Lemma 1 in
Cs\"{o}rg\H{o} et al (2003); see also Feller (1966).

\begin{lemma}
\label{csorgo} The following statements are equivalent:
\end{lemma}

\begin{enumerate}
\item $l(x)=EX^{2}I(|X|\leq x)$ is a slowly varying function at $\infty$;

\item $P(|X|>x)=o(x^{-2}l(x))$;

\item $E|X|I(|X|>x)=o(x^{-1}l(x))$;

\item $E|X|^{\alpha}I(|X|\leq x)=o(x^{\alpha-2}l(x))$ for $\alpha>2$.
\end{enumerate}

To clarify the behavior of the sequence of normalizer $B_{n}^{2}$ defined by
(\ref{defBn}) we state the following lemma that follows from relations (3.33)
and (3.44) in Kuelbs (1985).

\begin{lemma}
Assume (\ref{DA}) and define $\eta_{n}$ by (\ref{defeta}). Then, $l_{n}%
=l(\eta_{n})$ is a slowly varying function at $\infty$.
\end{lemma}

The next lemma is useful to study the variance of partial sums for truncated
random variables.

\begin{lemma}
\label{coeff}Under conditions of Theorem \ref{clt} and with the notation
(\ref{defcalpha}) and (\ref{defbn}) we have
\end{lemma}

\begin{enumerate}
\item The coefficients have the following order of magnitude: There are
constants $C_{1}$ and $C_{2}$ such that for all $n\geq1,$%
\begin{equation}
|b_{ni}|\leq C_{1}i^{1-\alpha}|L(i)|\text{ for }i\leq2n\text{ and }%
|b_{ni}|\leq C_{2}n(i-n)^{-\alpha}|L(i)|\text{ for }i>2n\ \text{.}%
\label{order}%
\end{equation}%
\begin{equation}
\sum_{i=1}^{\infty}b_{ni}^{2}\mathbb{\sim}c_{\alpha}n^{3-2\alpha}%
L^{2}(n)\text{ .}\label{sumb}%
\end{equation}

\item The asymptotic equivalence for the variance:
\begin{equation}
\sum_{i\geq1}b_{ni}^{2}l_{i}\mathbb{\sim}l_{n}\sum_{i\geq1}b_{ni}%
^{2}\mathbb{\sim}B_{n}^{2}\text{ ,} \label{VAR}%
\end{equation}
where $B_{n}^{2}$ is defined by (\ref{defBn}).

\item For any $p\geq1$ and any function $h(x)$ slowly varying at $\infty,$%
\begin{equation}
\sum_{i\geq1}|b_{ni}|^{p}i^{-1+p/2}|h(i)|\ll h(n)n^{p(3/2-\alpha)}%
L^{p}(n)\text{ .} \label{boundb}%
\end{equation}

\end{enumerate}

\textbf{Proof}. The fact that $|b_{ni}|\leq C_{1}i^{1-\alpha}|L(i)|$ for
$i\leq2n$ follows easily by the properties of slowly varying functions listed
in Lemma \ref{Karamata}.

For $i>2n$, by the properties of strong slowly varying functions, for $n$
sufficiently large:%
\[
(i-n)^{-\alpha}L(i-n)+...+i^{-\alpha}L(i)\leq\lbrack(i-n)^{-\alpha
}+...+i^{-\alpha}]\max_{i-n\leq j\leq i}L(j)\text{ .}%
\]
Then,
\[
\max_{i-n\leq j\leq i}L(j)\leq\max_{i/2\leq j\leq i}L(j)\ll L(i)
\]
since
\begin{equation}
\frac{\max_{m\leq j\leq2m}L(j)}{\min_{m\leq j<2m}L(j)}\rightarrow1\text{ .}
\label{propL}%
\end{equation}
The asymptotic equivalence in (\ref{sumb}) is well known. See for instance
Theorem 2 in Wu and Min (2005).

We turn now to show (\ref{VAR}). Let $M$ be a positive integer. We divide the
sum in $3$ parts, one from $1$ to $n$, one from $n+1$ to $nM,$ and the third
one with all the other terms. The idea of the proof is that for $n$ and $M$
large, the sum from $1$ to $nM$ dominates the sum of the rest of the terms.

We treat each of these three sums separately.

By using the definition of $b_{ni}=a_{1}+...+a_{i}=b_{i}$ for $1\leq i\leq n$
by analogy with the point 3 in Lemma \ref{Karamata} we show that%
\begin{equation}
\sum_{i=1}^{n}b_{i}^{2}l_{i}\mathbb{\sim}l_{n}\sum_{i=1}^{n}b_{i}^{2}\text{
.}\label{rel1}%
\end{equation}
To see this, by the first part of Lemma (\ref{Karamata}) we have $l_{n}%
=g_{n}h_{n}$ where $h_{n}=\exp\left(  \int_{B}^{n}\frac{a(y)}{y}dy\right)  $,
$g_{n}\rightarrow c>0$, $a(x)\rightarrow0$ as $x\rightarrow\infty,$ and $a(x)$
is continuous. It is easy to show that
\begin{equation}
h_{n}-h_{n-1}=o(h_{n}/n)\text{ as }n\rightarrow\infty\text{ }\label{propl}%
\end{equation}
and also, by the part 3 of the same lemma, we get $\sum_{i=1}^{n-1}b_{i}%
^{2}\ll nb_{n}^{2}$ .

Next, we have just to use the well known Toeplitz lemma:%
\begin{equation}
\lim_{n\rightarrow\infty}\frac{c_{n}}{d_{n}}=\lim_{n\rightarrow\infty}%
\frac{c_{n}-c_{n-1}}{d_{n}-d_{n-1}}\text{ ,} \label{toeplitz}%
\end{equation}
provided $d_{n}\rightarrow\infty$ and the limit in the right hand side exists.
Then, it follows that
\[
\lim_{n\rightarrow\infty}\frac{\sum_{i=1}^{n}b_{i}^{2}l_{i}}{l_{n}\sum
_{i=1}^{n}b_{i}^{2}}=\lim_{n\rightarrow\infty}\frac{\sum_{i=1}^{n}b_{i}%
^{2}l_{i}}{ch_{n}\sum_{i=1}^{n}b_{i}^{2}}=\lim_{n\rightarrow\infty}\frac
{b_{n}^{2}h_{n}}{h_{n}\sum_{i=1}^{n}b_{i}^{2}-h_{n-1}\sum_{i=1}^{n-1}b_{i}%
^{2}}\ \text{.}%
\]
We shall show that the limit in the right hand side is equal to $1$. We start
by writing
\[
h_{n}\sum_{i=1}^{n}b_{i}^{2}-h_{n-1}\sum_{i=1}^{n-1}b_{i}^{2}=(h_{n}%
-h_{n-1})\sum_{i=1}^{n-1}b_{i}^{2}+b_{n}^{2}h_{n}\text{ .}%
\]
Then, by (\ref{propl})%
\[
\lim_{n\rightarrow\infty}\frac{b_{n}^{2}h_{n}}{h_{n}\sum_{i=1}^{n}b_{i}%
^{2}-h_{n-1}\sum_{i=1}^{n-1}b_{i}^{2}}=\lim_{n\rightarrow\infty}\frac
{b_{n}^{2}h_{n}}{o(h_{n}/n)\sum_{i=1}^{n}b_{i}^{2}+b_{n}^{2}h_{n}}=1\text{ ,}%
\]
and (\ref{rel1}) follows.

To treat the second sum, notice that $l_{n}$ is increasing and then
\begin{equation}
\sum_{i=n+1}^{nM}b_{ni}^{2}l_{i}\mathbb{\sim}l_{n}\sum_{i=n+1}^{nM}b_{ni}^{2}
\label{rel2}%
\end{equation}
because
\[
l_{n}\sum_{i=n+1}^{nM}b_{ni}^{2}\leq\sum_{i=n+1}^{nM}b_{ni}^{2}l_{i}\leq
l_{nM}\sum_{i=n+1}^{nM}b_{ni}^{2}\text{ ,}%
\]
and $l_{n}$ is a function slowly varying at $\infty$ .

We treat now the last sum. By (\ref{sumb}), and Lemma \ref{Karamata}
\[
\sum_{i=nM+1}^{\infty}b_{ni}^{2}l_{i}\ll n^{2}\sum_{i=nM+1}^{\infty
}(i-n)^{-2\alpha}L^{2}(i)l_{i}\ll n^{2}[n(M-1)]^{1-2\alpha}L^{2}%
[nM]l_{nM}\text{ .}%
\]
We obtain%
\begin{equation}
\sum_{i=nM+1}^{\infty}b_{ni}^{2}l_{i}\ll B_{n}^{2}M^{1-2\alpha}\text{ as
}n\rightarrow\infty\text{ .} \label{rel3}%
\end{equation}
We combine now the estimates in (\ref{rel1}) and (\ref{rel2}). For $\delta>0$
fixed and $n$ sufficiently large
\[
(1-\delta)l_{n}\sum_{i\geq1}^{nM}b_{ni}^{2}\leq\sum_{i\geq1}^{nM}b_{ni}%
^{2}l_{i}\leq(1+\delta)l_{n}\sum_{i\geq1}^{nM}b_{ni}^{2}\text{ .}%
\]
Therefore,%
\begin{equation}
(1-\delta)l_{n}(\sum_{i\geq1}^{\infty}b_{ni}^{2}-\sum_{i>nM}b_{ni}^{2}%
)\leq\sum_{i\geq1}^{\infty}b_{ni}^{2}l_{i}\leq(1+\delta)l_{n}\sum_{i\geq
1}^{nM}b_{ni}^{2}+\sum_{i>nM}b_{ni}^{2}l_{i}\text{ .} \label{IneqB}%
\end{equation}
Then, by (\ref{rel3}), for a positive constant $C_{1}$ we have
\[
\lim\sup_{n\rightarrow\infty}\frac{1}{B_{n}^{2}}\sum_{i>nM}b_{ni}^{2}l_{i}%
\leq\frac{C_{1}}{M^{2\alpha-1}}\text{ .}%
\]
We also know that for a certain positive constant $C_{2}$,
\[
\lim\sup_{n\rightarrow\infty}\frac{1}{B_{n}^{2}}\sum_{i>nM}b_{ni}^{2}\leq
\frac{C_{2}}{M^{2\alpha-1}}\text{ .}%
\]
The result follows by dividing (\ref{IneqB}) by $B_{n}^{2}$ and taking first
$\lim\sup$ and also $\lim\inf$ when $n\rightarrow\infty$ followed by
$M\rightarrow\infty$ , and finally we let $\delta\rightarrow0$.

The proof of (\ref{boundb}) is similar and it is sufficient to divide the sum
in only two parts, one from $1$ to $2n$ and the rest. More exactly by using
(\ref{order}),
\[
\sum_{i=1}^{2n}|b_{ni}|^{p}i^{-1+p/2}|h(i)|\ll\sum_{i=1}^{2n}i^{p(1-\alpha
)}|L(i)|^{p}i^{-1+p/2}|h(i)|\ll h(n)n^{p(3/2-\alpha)}L^{p}(n)
\]
and,
\[
\sum_{i\geq2n}|b_{ni}|^{p}i^{-1+p/2}|h(i)|\ll n^{p}\sum_{i\geq2n}%
i^{-1+(1/2-\alpha)p}|L(i)|^{p}|h(i)|\ll h(n)n^{p(3/2-\alpha)}L^{p}(n)\text{ .}%
\]
The proof is complete. $\lozenge$

Next lemma is a variant of Theorem 12.3 in Billingsley (1968).

\begin{lemma}
\label{Ltight} Assume that $(X_{nk})_{1\leq k\leq n}$ is a triangular array of
centered random variables with finite second moment. For $0\leq m\leq n$ let
$S_{m}=\sum_{j=1}^{m}X_{nj}$ and for $0\leq t\leq1,$ $W_{n}(t)=S_{[nt]}.$
Assume that for every $\varepsilon>0$%
\begin{equation}
P(\max_{1\leq i\leq n}|X_{ni}|>\varepsilon)\rightarrow0 \label{condtight1}%
\end{equation}
and there is a positive constant $K,$ and an integer $N_{0}$ such that for any
$1\leq p<q\leq n$ with $q-p>N_{0}$ we have
\begin{equation}
E(S_{nq}-S_{np})^{2}\leq K(\frac{q}{n}-\frac{p}{n})^{\gamma}
\label{condtight2}%
\end{equation}
for some $\gamma>1.$ Then $W_{n}(t)$ is tight in $D[0,1]$, endowed with
Skorohod topology.
\end{lemma}

\textbf{Proof. }We shall base our proof on a blocking argument. We divide the
variables in blocks of size $N_{0}.$ Let $k=[n/N_{0}].$ For $1\leq j\leq k$
denote $Y_{nj}=\sum_{i=(j-1)N_{0}+1}^{jN_{0}}X_{ni}$ and $Y_{n,k+1}%
=\sum_{i=kN_{0}+1}^{n}X_{ni}.$ Define $V_{n}(t)=\sum_{j=1}^{[kt]}Y_{nk}$ .

Then we notice that it is enough to show that $V_{n}(t)$ is tight in $D[0,1]$
because by the fact that $[nt]-[kt]\leq2N_{0}$ and by (\ref{condtight1})
\[
P(\sup_{t}|W_{n}(t)-V_{n}(t)|>\varepsilon)\leq P(\max_{1\leq i\leq n}%
|X_{ni}|>\varepsilon/2N_{0})\rightarrow0\text{ .}%
\]
By Theorem 8.3 in Billingsley (1968) formulated for random elements of D (see
page 137 in Billingsley, 1968) we have to show that for every $0\leq t\leq1$
and $\varepsilon>0$ fixed,%

\[
\lim_{\delta\searrow0}\lim\sup_{n\rightarrow\infty}\frac{1}{\delta}%
P(\max_{[kt]\leq j\leq\lbrack k(t+\delta)]}|\sum_{i=[kt]}^{j}Y_{ni}%
|\geq\varepsilon)=0\text{ .}%
\]
By Theorem 12.2 in Billingsley (1968), because $\gamma>1,$ there is a constant
$K$ such that
\[
P(\max_{[kt]\leq j\leq\lbrack k(t+\delta)]}|\sum_{i=[kt]}^{j}Y_{ni}%
|\geq\varepsilon)\leq\frac{K}{\varepsilon^{2}}([\frac{k(t+\delta)N_{0}}%
{n}]-[\frac{ktN_{0}}{n}])^{\gamma}%
\]
and the result follows by multiplying with $1/\delta$ and passing to the limit
with $n\rightarrow\infty$ and then with $\delta\rightarrow0$. $\lozenge$

\section{Acknowledgement}

The authors are grateful to the referees for carefully reading the paper and
for numerous suggestions that significantly improved the presentation of the paper.

\end{document}